%% file: main.tex
\documentclass[runningheads]{llncs}
\usepackage[T1]{fontenc}
\usepackage{graphicx}
\usepackage{siunitx}
\usepackage{subcaption}
\usepackage{amsmath}
\usepackage{booktabs}
\usepackage{multirow}
\usepackage[ruled,vlined]{algorithm2e}
\newcommand{\etal}{\emph{et al.} }

\usepackage{paralist}
\usepackage{enumitem}
\usepackage{tikz,xcolor,hyperref}
\definecolor{lime}{HTML}{A6CE39}
\DeclareRobustCommand{\orcidicon}{%
	\begin{tikzpicture}
	\draw[lime, fill=lime] (0,0) 
	circle [radius=0.16] 
	node[white] {{\fontfamily{qag}\selectfont \tiny ID}};
	\draw[white, fill=white] (-0.0625,0.095) 
	circle [radius=0.007];
	\end{tikzpicture}
	\hspace{-2mm}
}
\foreach \x in {A, ..., Z}{%
	\expandafter\xdef\csname orcid\x\endcsname{\noexpand\href{https://orcid.org/\csname orcidauthor\x\endcsname}{\noexpand\orcidicon}}
}

\begin{document}
\title{LMEdge: QoS-Aware LLM Inference Orchestration on Edge Clusters \vspace{-.7cm}}
\author{Reza Farahani\inst{1}\orcidA{}
Zoha Azimi\inst{2}\orcidB{}
Mario Colosi\inst{3}\orcidC{} 
Schahram Dustdar\inst{1}\orcidD{}
\vspace{-.2cm}
}
\authorrunning{Reza Farahani et al.}
\titlerunning{LMEdge: QoS-Aware LLM Inference Orchestration on Edge Clusters}
\institute{Vienna University of Technology, Vienna, Austria \and
University of Klagenfurt, Klagenfurt, Austria\and
University of Messina, Messina, Italy
} 
\maketitle              
%%%%%%%%%%%%%%%%%%%Abstract%%%%%%%%%%%%%%%%%%%
\vspace{-.8 cm}
\begin{abstract}
Large language model (LLM) services increasingly operate on edge infrastructure, enabling low-latency and privacy-preserving AI services. However, efficiently serving LLM requests across heterogeneous and resource-constrained edge devices requires orchestration mechanisms that jointly determine model configuration (family, size, and quantization level) and execution placement while satisfying user- and system-level quality of service (QoS) requirements. This paper introduces \texttt{LMEdge}, a QoS-aware orchestration service that dynamically makes these decisions across heterogeneous edge devices. We formulate the problem as a binary integer linear programming (BILP) optimization that minimizes response time under accuracy, network, and resource constraints. To enable scalable online scheduling, we employ five lightweight machine learning (ML) models to predict query-specific latency, accuracy, resource usage, and response size for each model–size-quantization-device combination, and design a lightweight heuristic that approximates the BILP solution. We collect a comprehensive benchmarking dataset of over \num{59000} rows to train models and support reproducibility. Evaluation on a Kubernetes-based edge testbed with \num{57} instances and diverse query categories shows that \texttt{LMEdge} reduces latency, preserves accuracy, improves resource utilization, and increases serving ratio compared to two baselines.
\vspace{-.3cm}
\end{abstract}
\vspace{-1cm}
\input{01_Introduction}
\input{02_RelatedWork}
\input{04_Model}
\input{05_Arch}

\input{06_Alg}
\input{07_Evaluation}

\input{08_Experiments}

\input{09_Conclusion}
%%%%%%
\vspace{-.4cm}
\bibliographystyle{./Bibliography/splncs04}
\bibliography{./Bibliography/main.bib}
\end{document}

%% file: 01_Introduction.tex
\vspace{-.2cm}
\section{Introduction}
\label{sec:Introduction }
\vspace{-.2cm}
Recently, LLMs have become the backbone of AI services like chatbots, assistants, and code generators, with unprecedented adoption exemplified by ChatGPT reaching \num{100} million users in two months~\cite{Demandsage}.  While closed-source foundation models such as GPT are typically deployed in cloud environments, open-source LLMs are increasingly being served on edge infrastructure to provide lower latency, enhanced privacy, and reduced reliance on centralized resources. These deployments comprise heterogeneous devices with diverse computational, memory, and networking capabilities while simultaneously serving queries with varying complexity, length, and quality requirements. This heterogeneity raises a fundamental systems challenge: \emph{``How can queries be efficiently served under constrained and dynamically changing resource conditions while maintaining target latency and accuracy?''} However, existing approaches such as split execution~\cite{patel2024splitwise} and edge model caching~\cite{xu2025serving} provide only coarse-grained optimization. They typically focus on a single dimension of the deployment space and lack per-query orchestration mechanisms that jointly select the LLM \emph{family}, \emph{size}, \emph{quantization level}, and \emph{execution device} to satisfy both user QoS requirements (e.g., latency and accuracy) and system-level constraints (e.g., resource availability, computational load, and network bandwidth).

This paper presents \texttt{LMEdge}, a QoS-aware orchestration service that dynamically selects the LLM family, model size, quantization level, and execution device for each query across heterogeneous edge devices. We formulate query orchestration as a BILP optimization model to minimize the query response time, considering accuracy and system constraints. We design a modular architecture and a lightweight online scheduling heuristic that leverages five ML models to predict inference latency, response accuracy and size, and resource use (i.e., CPU/GPU and memory). We design a realistic Kubernetes-based edge testbed with \num{57} instances with \num{29} model instances across families, sizes, and quantization levels. We build a benchmarking dataset of over \num{59000} query–model–quantization–device records to train predictors and support reproducibility using more than \num{1422} user queries from five public datasets spanning seven categories (e.g., mathematics, programming, science, and commonsense reasoning). Experimental results confirm that \texttt{LMEdge} achieves better latency, accuracy, resource utilization, and serving ratios compared to two baselines.
%The paper has eight sections. Section~\ref{sec:RelatedWork} reviews the related work before formulating the problem in Section~\ref{sec:system}. Section~\ref{sec:arch} presents the LMEdge system architecture before explaining its heuristics in Section~\ref{sec:alg}. Section~\ref{sec:setup} describes the setup and Section~\ref{sec:result} analyzes the results. Finally, Section~\ref{sec:conc} concludes the paper.

%% file: 02_RelatedWork.tex
\vspace{-.3 cm}
\section{Related Work}
\label{sec:RelatedWork}
\vspace{-.3 cm}
Oh~\etal proposed ExeGPT~\cite{oh2024exegpt}, an LLM scheduler that leverages input/output length distributions to optimize resource allocation, batch sizing, and parallelism. Stojkovic~\etal introduced DynamoLLM~\cite{stojkovic2025dynamollm}, an energy‑aware framework that dynamically adjusts GPU resources (i.e., parallelism, frequencies) to reduce energy consumption and satisfy latency constraints. 
Dai~\etal proposed C2MAB‑V~\cite{dai2024cost}, a cost‑aware multi‑armed bandit algorithm that learns optimal LLM combinations across cloud and local deployments, balancing query quality and inference cost. 
Liu~\etal introduced OptLLM~\cite{liu2024optllm}, an offline framework that predicts LLM performance and generates Pareto‑optimal query allocations, minimizing cost or maximizing accuracy under given budget and performance goals.
Ong~\etal introduced RouteLLM~\cite{ong2024routellm}, a learning scheme trained on human preference data to direct queries between LLM instances, cutting inference costs while preserving user‑aligned output quality.
Yao~\etal used diffusion-based reinforcement learning for coordinating edge–cloud LLM query execution~\cite{yao2025enhancing}. 
Edge-LLM~\cite{cai2024edge} and EdgeShard~\cite{zhang2024edgeshard} applied model partitioning and device selection on the edge-cloud without QoS guarantees.
Existing approaches optimize primarily for latency and overlook heterogeneity across LLMs, quantization levels, and devices. They also lack predictive scheduling mechanisms. To address these gaps, \texttt{LMEdge}: \emph{(i)} performs QoS-aware per-query orchestration; \emph{(ii)} employs ML predictors for query-specific performance estimation; \emph{(iii)} combines BILP optimization with a lightweight heuristic; \emph{(iv)} provides a benchmark of over \num{59000} query-model-quantization-device measurements; and \emph{(v)} is validated on a heterogeneous Kubernetes edge testbed using thousands of real queries.

%% file: 04_Model.tex
\section{Problem Formulation}\label{sec:system}
\vspace{-.3cm}
We denote $\mathcal{Q}$ as a continuous stream of queries, each with an arrival time $t_{q}$. At scheduling epoch $t$, \texttt{LMEdge} considers the set of newly arrived queries $\mathcal{Q}_t$ together with deferred queries from epoch $\mathcal{D}_{t-1}$ for processing.  
We represent each query $q\in\mathcal{Q}$ by an input feature vector $\phi_{q}$, derived from the prompt (e.g., length, lexical diversity). We also define the edge cluster as a hierarchical pool of heterogeneous compute resources  
$\mathcal{I}=\{\mathcal{I_{RR}}\cup\mathcal{I_{RC}}$\},  
where $\mathcal{I_{RR}}$ and $\mathcal{I_{RC}}$ denote the sets of resource-rich (ERR) and -constrained (ERC) edge instances.
At each scheduling epoch $t$, \texttt{LMEdge} monitors the following metrics for each instance $i\in\mathcal{I}$:
\textit{ (i) Available computation} $\Omega_{i}$, including the number of free CPU/GPU units ($\Omega_{i}^{Cmp}$) and memory in GB ($\Omega_{i}^{Mem}$); 
\textit{ (ii) Available bandwidth} $\mu_{i}$, denoting the currently available  uplink/downlink capacity between edge instance $i$ and \texttt{LMEdge}, which constrains prompt upload and response delivery for queries assigned to $i$; and 
\textit{ (iii) Concurrency limit} $\rho_{i}$, the maximum number of inferences that instance $i$ can execute in parallel without violating its service-level guarantees, abstracting provider quotas. 
Let $\mathcal{M}$ denote the set of deployed models, including both 
LLMs with multi-billion parameters. Each model $m\in\mathcal{M}$ has: 
\textit{(i) Quantization levels} $\mathcal{C}^m$, denoting the set of precision formats (e.g., Q4, Q8, FP16) supported by model $m\in\mathcal{M}$; and 
\textit{(ii) Deployment status} $\alpha:\mathcal{M} \times\mathcal{C}\times\mathcal{I}\to\{0,1\}$, where 
$\alpha(m,c,i)=1$ indicates that compute instance $i\in \mathcal{I}$ hosts model $m$ at quantization $c$, 
and $\alpha(m,c,i)=0$ otherwise.

%%%%%%

We use ML-based models to predict query-specific behavior across model, quantization, and instance combinations. Each predictor takes the query features from the vector $\phi_q$, along with model-instance details. These models are:
\textit{(i) Inference time} $\mathcal{F}_{q,i}^{m,c}$, predicts inference time for serving $q$ using model $m$ at quantization $c$ on instance $i$; 
\textit{(ii) Accuracy} $\mathcal{A}_{q,i}^{m,c}$, predicts the response accuracy of model $m$ at quantization $c$ when applied to $q$ on instance $i$; 
\textit{(iii) Compute demand} $\mathcal{Z}_{q,i}^{m,c}$, predicts CPU/GPU demands for serving $q$ with model $m$ at quantization $c$ on instance $i$;
\textit{(iv) Memory demand} $\mathcal{R}_{q,i}^{m,c}$, predicts memory demands for serving $q$ with model $m$ at quantization $c$ on instance $i$; and 
\textit{(v) Response size} $\mathcal{L}_{q,i}^{m,c}$, predicts response size for $q$ using model $m$ with quantization $c$ on instance $i$.
We also define the serving time of each query $q\in\mathcal{Q}$ as the sum of model inference delay and communication delay. We use a binary assignment variable $\mathcal{B}^{m,c}_{q,i}\in\{0,1\}$ to indicate whether query $q$ is assigned to model $m \in \mathcal{M}$ at quantization level $c \in\mathcal{C}^m$ deployed on instance $i\in\mathcal{I}$. 
Thus, in Eq.~(\ref{eq:1}) ensures that each query is mapped at most to one model-quantization-instance triple:
\begin{equation}
\label{eq:1}
\sum_{m\in\mathcal{M}} \sum_{c\in\mathcal{C}^m}\sum_{i\in\mathcal{I}} \mathcal{B}^{m,c}_{q,i}\cdot\alpha(m,c,i) \leq 1, \qquad \forall q\in\{\mathcal{Q}_{t}\cup\mathcal{D}_{t-1}\}.
\end{equation}
We define the total response time $T^{m,c}_{q,i}$ of query $q$ serving by model $m$ with quantization $c$ on instance $i$ as:
\begin{equation}
\label{eq:2}
T^{m,c}_{q,i} = \underbrace{\mathcal{F}^{m,c}_{q,i}}_\text{Inference delay} + \underbrace{\underbrace{\frac{(\delta^{in}_q + \mathcal{L}^{m,c}_{q,i})}{\mu_i}}_\text{Transmission delay}\cdot \underbrace{(1+\lambda \cdot \frac{l_i}{\rho_i})} _\text{Congestion control}}_\text{Congestion-aware delay}.
\end{equation}
where $\delta^{in}_q$ is the input size of query $q$, $l_i$ is the number of active inferences on $i$, and $\lambda$ is a tunable congestion penalty. While $\mathcal{F}^{m,c}_{q,i}$ predicts model-specific computation delay, it omits queuing and serialization overheads under fluctuating load. The congestion-aware multiplier augments the transmission term to capture contention, enabling \texttt{LMEdge} to favor less‑loaded instances and yielding less end‑to‑end latency.
We define the best achievable accuracy of $q\in\mathcal{Q}$ as the maximum predicted accuracy across all model–quantization–instance combinations:  
%%%
\begin{equation}
\label{eq:3}
a^{\text{best}}_q = \max_{m,c,i} \mathcal{A}^{m,c}_{q,i}.
\end{equation}
%%%
We then enforce that the selected assignment $(m,c,i)$ for query $q$ achieves accuracy within a tolerated fraction of $a^{\text{best}}_q$: 
\begin{equation}
\label{eq:3-tolerated}
\mathcal{A}^{m,c}_{q,i} \geq (1 - \theta)\cdot a^{\text{best}}_q \quad \forall q \in \{\mathcal{Q}_{t} \cup \mathcal{D}_{t-1}\}~\text{and}~\mathcal{B}_{q,i}^{m,c}=1
\end{equation}
where \mbox{$\theta\in[0,1]$} bounds the permissible accuracy loss, with small values enforcing near‑optimal accuracy and larger values allowing lower‑accuracy but faster or more resource‑efficient assignments. For example, if \mbox{$a^{\text{best}}_q=90\%$} and $\theta=0.1$, any assignment with at least \qty{81}{\percent} accuracy is acceptable, enabling \texttt{LMEdge} to trade a \qty{9}{\percent} accuracy reduction for better latency. 
%%%%%%%
To guarantee feasibility, query assignments at epoch $t$ must not exceed the available capacity of any compute instance $i \in \mathcal{I}$ across compute, memory, bandwidth, and concurrency dimensions. In addition, Eqs.~(\ref{eq:4})--(\ref{eq:5}) ensure that the total computational demands of scheduled queries in $\mathcal{Q}_t \cup \mathcal{D}_{t-1}$ do not exceed the available resources of $i\in\mathcal{I}$:
\begin{equation}
\label{eq:4}
\sum_{q\in\{\mathcal{Q}_t\cup\mathcal{D}_{t-1}\}}\sum_{m\in \mathcal{M}}\sum_{c\in\mathcal{C}^m}
\mathcal{B}_{q,i}^{m,c}\cdot \mathcal{Z}^{m,c}_{q,i}\leq \Omega_{i}^{\text{Cmp}},
\end{equation}
\begin{equation}
\label{eq:5}
\sum_{q\in\{\mathcal{Q}_t\cup\mathcal{D}_{t-1}\}}\sum_{m\in \mathcal{M}}\sum_{c\in\mathcal{C}^m}
\mathcal{B}_{q,i}^{m,c}\cdot \mathcal{R}^{m,c}_{q,i}\leq \Omega_{i}^{\text{Mem}}.
\end{equation}

Moreover, Eq.~(\ref{eq:7}) ensures that the cumulative input and output sizes of queries assigned to $i\in\mathcal{I}$ do not exceed its available bandwidth $\mu_i$:
\begin{equation}
\label{eq:7}
\sum_{q\in\{\mathcal{Q}_t\cup\mathcal{D}_{t-1}\}}\sum_{m \in \mathcal{M}} \sum_{c \in \mathcal{C}^m} 
\mathcal{B}^{m,c}_{q,i} \cdot (\delta^{in}_q + \mathcal{L}^{m,c}_{q,i})\leq\mu_i.
\end{equation}
%%%
In addition, Eq.~(\ref{eq:8}) ensures that the sum of already running inferences $l_i$ and newly scheduled assignments does not exceed the concurrency threshold $\rho_i$, ensuring that provider-imposed concurrency quotas are never violated. 
\begin{equation}
\label{eq:8}
\sum_{q\in\{\mathcal{Q}_t\cup\mathcal{D}_{t-1}\}}\sum_{m\in \mathcal{M}}\sum_{c\in\mathcal{C}^m}
\mathcal{B}^{m,c}_{q,i}\leq\rho_i - l_i.
\end{equation}
%%%%%%%

Finally, the BILP model (Eq.~\ref{eq:9}) assigns each query \mbox{$q\in\{\mathcal{Q}_t\cup\mathcal{D}_{t-1}\}$} to a model–quantization–instance triple, minimizing the total response time:

%%%%%%%%
\begin{equation}
\label{eq:9}
\begin{aligned}
\textit{Minimize}\quad 
& \sum_{q\in\{\mathcal{Q}_t\cup\mathcal{D}_{t-1}\}} 
  \sum_{m\in \mathcal{M}}
  \sum_{c\in\mathcal{C}^m}
  \sum_{i\in \mathcal{I}}
  \mathcal{B}^{m,c}_{q,i} \cdot T^{m,c}_{q,i}, \\[0.3em]
\text{s.t.}\quad 
& \text{constraints Eqs.~(\ref{eq:1})–(\ref{eq:8})}, 
\text{vars.}
& \mathcal{B}_{q,i}^{m,c} \in \{0,1\}.
\end{aligned}
\end{equation}
%%%%%%
Queries that cannot be placed within defined constraints are deferred into $\mathcal{D}_t$ for reconsideration in epoch \mbox{$t+1$}. 
%The problem is a multi‑dimensional multiple‑choice NP‑hard~\cite{garey2002computers}, necessitating lightweight scheduling for practical scenarios.
%%%%%%%%%%%%%%%%%%%%%
\begin{figure}[!t]
	\centering
\vspace{-.5cm}
\includegraphics[width=1\linewidth]{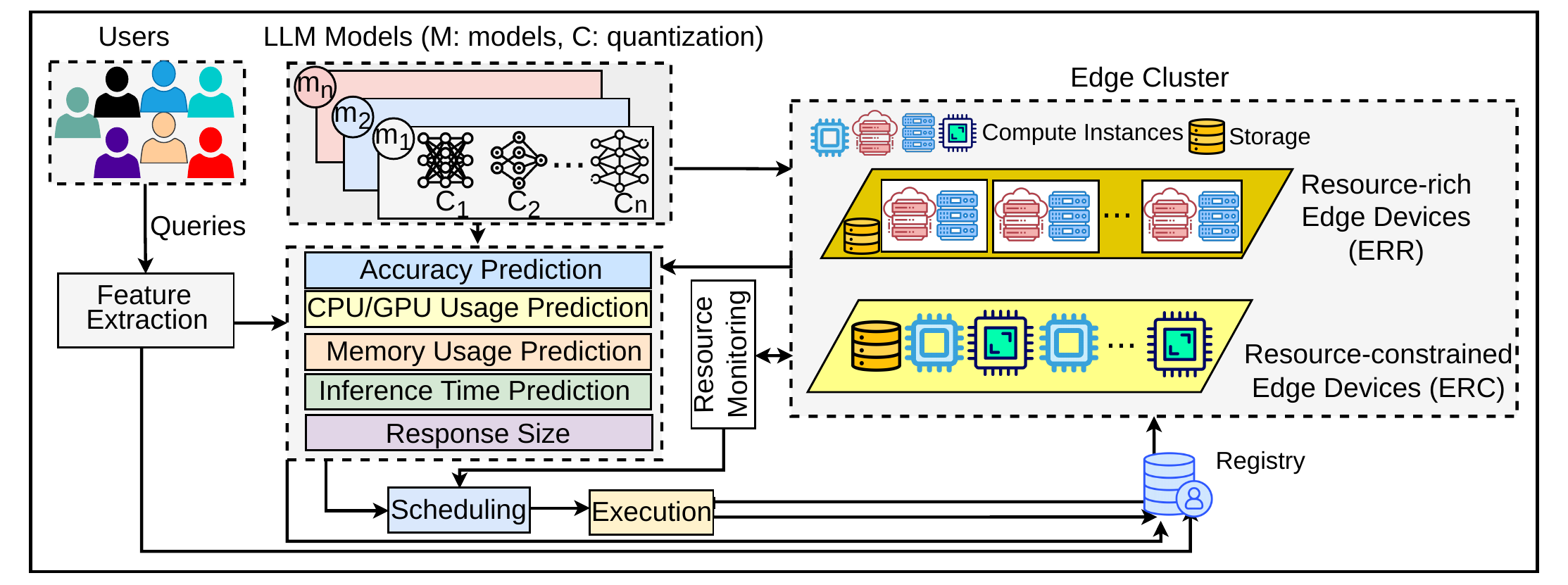}
	\vspace{-.5cm}
    \caption{LMEdge system architecture.}
	\label{arch}
    \vspace{-.5cm}
\end{figure}

%% file: 05_Arch.tex
\vspace{-.3cm}
\section{LMEdge System Architecture}
\label{sec:arch}
\vspace{-.3cm}
Fig.~\ref{arch} illustrates the modular \texttt{LMEdge} architecture. As a control-plane service, \texttt{LMEdge} orchestrates LLM queries across heterogeneous edge devices through six modules:
\textbf{1) Feature extraction} transforms each query into a feature vector comprising structural (e.g., prompt length, sentence count) and linguistic (e.g., lexical diversity, unique word count) features.
\textbf{2) Prediction} estimates query-specific accuracy, latency, resource usage, and response size for each model-quantization-instance combination.
\textbf{3) Resource monitoring} tracks resource availability, including CPU/GPU utilization, memory, bandwidth, and concurrency limits.
\textbf{4) Scheduling} combines predictions and resource states to select the model, quantization level, and execution placement that satisfy QoS and resource constraints while minimizing response time.
\textbf{5) Execution} invokes the selected model on the assigned instance and reports runtime observations for adaptation and retraining.
\textbf{6) Registry} stores query features, prediction outputs, and execution traces to support reproducibility, predictor refinement, and similarity-based reuse.

%% file: 06_Alg.tex
\vspace{-.3cm}
\section{LMEdge Heuristic Algorithm}
\label{sec:alg}
\vspace{-.3cm}
The \texttt{LMEdge} algorithm (Alg.~\ref{alg:heuristic}) efficiently approximates the NP-hard optimization problem in Eq.~(\ref{eq:9}) for scalable online scheduling. Given incoming and deferred queries, deployed LLMs and quantizations, available compute instances, QoS parameters, and monitoring data, it produces a scheduling plan $Sched_t$ and an updated deferred-query set $\mathcal{D}_t$. \texttt{LMEdge} follows a non-preemptive execution model and consists of three phases:
\textbf{1) Prediction:} for each query $q\in\{\mathcal{Q}_t \cup \mathcal{D}_{t-1}\}$, deployed model $m\in\mathcal{M}$, quantization $c\in\mathcal{C}^m$, and compute instance $i\in\mathcal{I}$, the algorithm first checks deployability via $\alpha(m,c,i)$ (lines 2--3). For feasible combinations, it predicts accuracy, latency, CPU/GPU utilization, memory usage, and response size, storing the results in $\mathcal{P}$ (line 5). It also records the best achievable accuracy $a_q^{best}$ for subsequent filtering (line 6). \textbf{2) Ranking and sorting:} the algorithm merges newly arrived and deferred queries (line 7), estimates the minimum achievable response time $\eta_q$ across all feasible assignments using Eq.~(\ref{eq:2}) (lines 8--13), and sorts queries in ascending order of $\eta_q$, breaking ties by arrival time (line 14). \textbf{3) Scheduling:} queries are processed in sorted order (line 15). For each query, feasible model-quantization-instance assignments are evaluated against accuracy and resource constraints (lines 17--22). The assignment with the minimum estimated response time is selected and added to $Sched_t$ (lines 23--25) and updates the selected resources (lines 26--27); otherwise, the query is deferred to $\mathcal{D}_t$ (lines 28--29). The algorithm finally returns $Sched_t$ and $\mathcal{D}_t$ (line 30).
Alg.~(\ref{alg:heuristic}) has a worst-case time complexity of $\mathcal{O}\left(|\mathcal{Q}_t| \cdot |\mathcal{M}| \cdot |\mathcal{C}^m| \cdot |\mathcal{I}|\right)$, covering prediction, ranking, and scheduling phases.
%%%%%%%%
\begin{algorithm}[!t]
\LinesNumbered
\fontsize{7pt}{7pt}\selectfont
\caption{LMEdge Scheduling Algorithm}
\label{alg:heuristic}

\KwIn{$\mathcal{Q}_t$, $\mathcal{D}_{t-1}$, $\mathcal{M}$, $\mathcal{C}^m$, $\mathcal{I}$, $\theta, \lambda$}
\KwOut{$Sched_t$, $\mathcal{D}_t$}

$\mathcal{P}_t \leftarrow \emptyset$, \quad $\mathcal{D}_t \leftarrow \emptyset$

\For{$q\in\{\mathcal{Q}_t \cup \mathcal{D}_{t-1}\}$}{
\For{$(m,c,i) \in \mathcal{M} \times \mathcal{C}^m \times \mathcal{I}$}{
    % \For{$m \in \mathcal{M}$}{
    %     \For{$c \in \mathcal{C}^m$}{
    %         \For{$i \in \mathcal{I}$}{
                \If{$\alpha(m,c,i) = 1$}{
                    $\mathcal{P}_t[q,m,c,i] \leftarrow \{
                        \mathcal{F}^{m,c}_{q,i},
                        \mathcal{A}^{m,c}_{q,i},
                        \mathcal{Z}^{m,c}_{q,i},
                        \mathcal{R}^{m,c}_{q,i},
                        \mathcal{L}^{m,c}_{q,i}
                    \}$\tcp{Prediction phase}
                }
            }
    %     }
    % }
    $a^{\text{best}}_q \leftarrow \max_{m,c,i} \mathcal{A}^{m,c}_{q,i}$
}

$\mathcal{Q}_t \leftarrow \{\mathcal{Q}_t \cup \mathcal{D}_{t-1}\}$

\For{$q \in \mathcal{Q}_t$}{
    $\eta_q \leftarrow \infty$
    
\For{$(m,c,i) \in \mathcal{M} \times \mathcal{C}^m \times \mathcal{I}$}{
    % \For{$m \in \mathcal{M}$}{
    %     \For{$c \in \mathcal{C}^m$}{
    %         \For{$i \in \mathcal{I}$}{
                \If{$\alpha(m,c,i) = 1$}{
                    $T^{m,c}_{q,i} \leftarrow \mathcal{F}^{m,c}_{q,i}
                    + \frac{(\delta^{in}_q + \mathcal{L}^{m,c}_{q,i})}{\mu_i}
                    \cdot \left(1 + \lambda \frac{l_i}{\rho_i}\right)$

                    $\eta_q \leftarrow \min(\eta_q, T^{m,c}_{q,i})$
                    \tcp{Ranking phase}
                }
            }
            
        }
%     }
% }

$\mathcal{Q}_t \leftarrow \text{Sort}(\mathcal{Q}_t, (\eta_q, t_q) \uparrow)$
\tcp{Sorting phase}

\For{$q \in \mathcal{Q}_t$}{
    $S_q \leftarrow \emptyset$
    
\For{$(m,c,i) \in \mathcal{M} \times \mathcal{C}^m \times \mathcal{I}$}{
    % \For{$m \in \mathcal{M}$}{
    %     \For{$c \in \mathcal{C}^m$}{
    %         \For{$i \in \mathcal{I}$}{
                \If{$\alpha(m,c,i) = 1$}{
                    $T^{m,c}_{q,i} \leftarrow \mathcal{F}^{m,c}_{q,i}
                    + \frac{(\delta^{in}_q + \mathcal{L}^{m,c}_{q,i})}{\mu_i}
                    \cdot \left(1 + \lambda \frac{l_i}{\rho_i}\right)$

                    \If{$\mathcal{A}^{m,c}_{q,i} \geq (1 - \theta)\cdot a^{\text{best}}_{q}$}{
                        \If{
                            $\mathcal{Z}^{m,c}_{q,i} \leq \Omega_i^{Cmp}$
                            \textbf{and}
                            $\mathcal{R}^{m,c}_{q,i} \leq \Omega_i^{Mem}$
                            \textbf{and}
                            $\delta^{in}_q + \mathcal{L}^{m,c}_{q,i} \leq \mu_i$
                            \textbf{and}
                            $l_i \leq \rho_i$
                        }{
                            $S_q \leftarrow S_q \cup \{(m,c,i,T^{m,c}_{q,i})\}$
                        }
                    }
                }
            }
        }
    % }

    \If{$S_q \neq \emptyset$}{
        $(m^*, c^*, i^*) \leftarrow \arg\min_{(m,c,i)\in S_q} T^{m,c}_{q,i}$
        
        $Sched_t \leftarrow Sched_t \cup (q,m^*,c^*,i^*)$
        \tcp{Scheduling phase}

        $\Omega_{i^*}^{Cmp} \leftarrow \Omega_{i^*}^{Cmp} - \mathcal{Z}^{m^*,c^*}_{q,i^*}$;
        $\Omega_{i^*}^{Mem} \leftarrow \Omega_{i^*}^{Mem} - \mathcal{R}^{m^*,c^*}_{q,i^*}$;
        $\mu_{i^*} \leftarrow \mu_{i^*} - (\delta^{in}_q + \mathcal{L}^{m^*,c^*}_{q,i^*})$;
        $l_{i^*} \leftarrow l_{i^*} + 1$
        
    }
    \Else{
        $\mathcal{D}_t \leftarrow \mathcal{D}_t \cup \{q\}$
    }
% }
\Return{$Sched_t$, $\mathcal{D}_t$}
\end{algorithm}
%%%%%%%%%

%% file: 07_Evaluation.tex
\vspace{-.8cm}
\section{Evaluation Setup}
\label{sec:setup}
\vspace{-.3cm}
We built a realistic edge testbed comprising two Kubernetes clusters with \num{12} virtual and \num{45} physical instances. Both clusters run \texttt{Kubernetes}~1.32.3 with \texttt{containerd}~1.7.24 and are interconnected via \texttt{Submariner} 0.20.0 (Globalnet mode). A minimal \texttt{Istio}~1.25.2 deployment handles query routing, while containerized \texttt{Ollama} backends serve multiple quantized LLMs. Model weights are stored in a centralized \texttt{MinIO} S3-compatible repository with \qty{1}{\tera\byte} capacity.
The ERR cluster consists of a master node (12 CPU cores, \SI{32}{GB} RAM) and \num{12} worker VMs running \texttt{Ubuntu}~22.04 LTS: \SI{2}{\times} large (8 CPU/\SI{32}{GB}), \SI{6}{\times} medium (4 CPU/\SI{24}{GB}), and \SI{4}{\times} small (2 CPU/\SI{16}{GB}). The ERC cluster comprises a master node (24 CPU cores, \SI{32}{GB} RAM), \num{37} Raspberry Pi devices, and \num{6} NVIDIA Jetson Nano devices, interconnected through two \texttt{TP-LINK} switches.
%%%
\begin{table}[!t]
\centering 
\caption{LLM models deployed across the devices.}
\label{tab:llm-depl}
\fontsize{7pt}{7pt}\selectfont
\begin{tabular}{c|c|c|c|}
\cline{2-4}
\multicolumn{1}{l|}{} & \textit{Model} & \textit{Params} & \textit{Quantization Level} \\ \hline
\multicolumn{1}{|c|}{\multirow{9}{*}{\textit{ERR VMs}}} & Gemma2 & \num{2}B & Q4\_0 \\ \cline{2-4}
\multicolumn{1}{|c|}{} & \multirow{2}{*}{Gemma3} & \num{1}B & $\text{Q}\{3,4\}\text{\_K\_}\{S,M\}$ \\ \cline{3-4}
\multicolumn{1}{|c|}{} & & \num{12}B & Q4\_K\_M \\ \cline{2-4}
\multicolumn{1}{|c|}{} & Llama3.1 & \num{8}B & Q8\_0, F16 \\ \cline{2-4}
\multicolumn{1}{|c|}{} & \multirow{2}{*}{Llama3.2} & \num{1}B & Q4\_K\_M, Q8\_0 \\ \cline{3-4}
\multicolumn{1}{|c|}{} & & \num{3}B & $\text{Q}\{2\text{-}4\}\text{\_K\_}\{S,M,L\}\text{, Q4\_0}$ \\ \cline{2-4}
\multicolumn{1}{|c|}{} & \multirow{2}{*}{Qwen3} & \num{0.6}B & $\text{Q4\_K\_}\{S,M\}\text{, Q8\_0, BF16}$ \\ \cline{3-4}
\multicolumn{1}{|c|}{} & & \num{1.7}B & Q4\_K\_M \\ \cline{2-4}
\multicolumn{1}{|c|}{} & Mistral & \num{7}B & Q8\_0 \\ \hline
\multicolumn{1}{|c|}{\multirow{3}{*}{\textit{ERC NJNs}}} & Gemma3 & \num{1}B & $\text{Q}\{3,4\}\text{\_K\_}\{S,M\}$ \\ \cline{2-4}
\multicolumn{1}{|c|}{} & Llama3.2 & \num{1}B & Q8\_0 \\ \cline{2-4}
\multicolumn{1}{|c|}{} & \multirow{2}{*}{TinyLlama} & \multirow{2}{*}{\num{1.1}B} & \multirow{2}{*}{$\text{Q}\{2\text{-}4\}\text{\_K\_}\{S,M,L\}\text{, Q}\{4,8\}\text{\_0}$} \\ \cline{1-1}
\multicolumn{1}{|c|}{\textit{ERC RPis}} & & & \\ \hline
\end{tabular}
\end{table}
%%%%%%%%%%%

\begin{figure}[!t]
\centering
\begin{subfigure}[b]{0.48\linewidth}
    \centering
    \includegraphics[width=\linewidth]{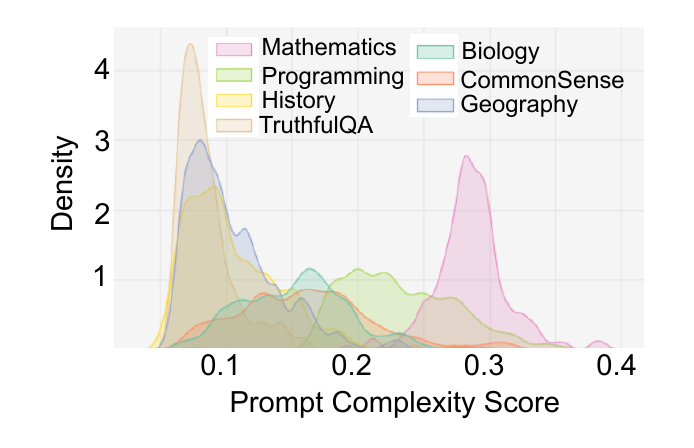}
    \vspace{-.5cm}
    \caption{Prompt complexity (KDE).}
    \label{dataset-complexity}
\end{subfigure}
\hfill
\begin{subfigure}[b]{0.48\linewidth}
    \centering
    \includegraphics[width=\linewidth]{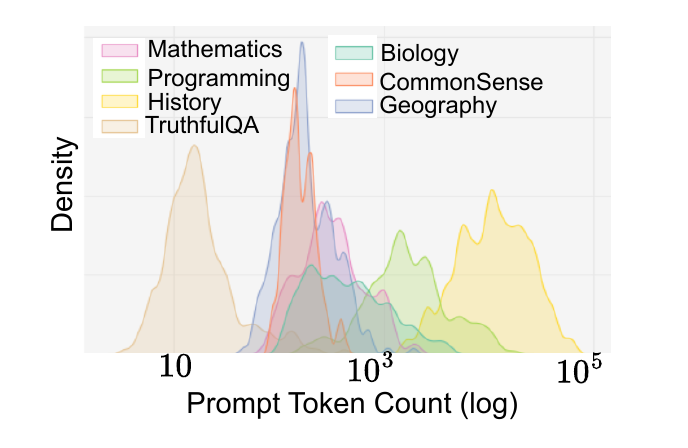}
    \vspace{-.5cm}
    \caption{Prompt token count (KDE).}
    \label{dataset-token}
\end{subfigure}
\vspace{2mm}
\begin{subfigure}[b]{0.48\linewidth}
    \centering
    \includegraphics[width=\linewidth]{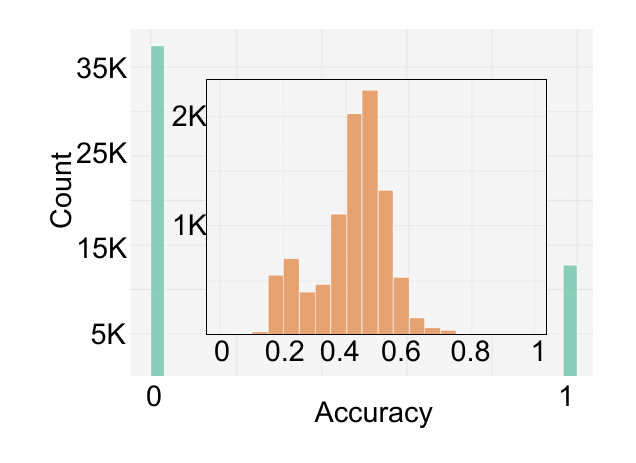}
    \vspace{-.5cm}
    \caption{Accuracy.}
    \label{dataset-accuracy}
\end{subfigure}
\hfill
\begin{subfigure}[b]{0.48\linewidth}
    \centering
    \includegraphics[width=\linewidth]{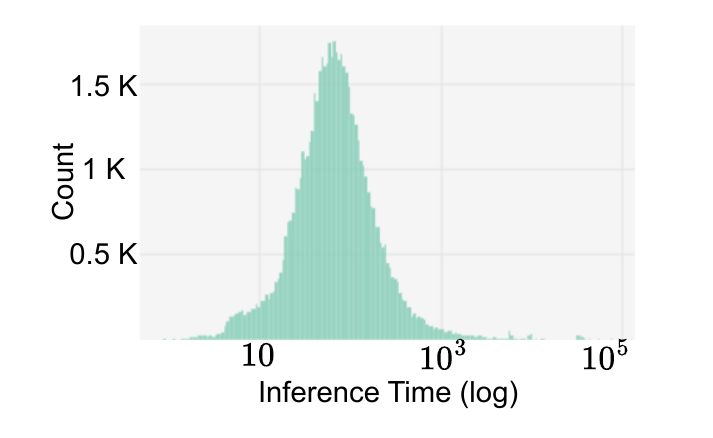}
    \vspace{-.5cm}
    \caption{Inference time (s).}
    \label{dataset-inf}
\end{subfigure}
\vspace{-.5cm}
\caption{Distribution of query complexity, accuracy, and inference time in the benchmarking dataset.}
\label{fig:benchmark}
\vspace{-.5cm}
\end{figure}

We deploy \num{29} LLM configurations across the ERR and ERC layers, spanning \num{8} open-source model families (e.g., Gemma2 and Llama 3.2) with sizes ranging from \num{0.6}B to \num{12}B parameters. Models are configured using \num{10} quantization formats alongside FP16 and BF16 precision. Resource-constrained ERC devices (Raspberry Pi and Jetson Nano) host lightweight quantized models, whereas ERR instances additionally serve larger models of up to \num{12}B parameters. Table~\ref{tab:llm-depl} summarizes the deployment.
We evaluate \texttt{LMEdge} using \num{1422} queries from five benchmark datasets covering mathematics, code generation, history, commonsense reasoning, and truthfulness. The workload comprises three \texttt{MMLU} subsets (biology, world history, and geography), \num{200} randomly sampled queries from each of \texttt{GSM8K}, \texttt{CommonsenseQA}, and \texttt{TruthfulQA}, and all \num{164} tasks from \texttt{HumanEval}. All datasets provide ground-truth answers, enabling accuracy assessment. We use the \texttt{Nvidia NeMoCurator} prompt complexity classifier~\cite{complexity} to assess query complexity based on factors such as reasoning, creativity, and required knowledge. Fig.~\ref{dataset-complexity} shows the kernel density estimation (KDE) of complexity scores, indicating that the \num{1422} queries span a broad range of difficulty levels, while Fig.~\ref{dataset-token} presents the KDE of prompt token counts, showing variation in query length.
Our benchmarking dataset comprises over \num{59000} measurements spanning diverse query, model, quantization, and device combinations. Each record captures CPU/GPU and memory utilization, inference latency, response size, and accuracy. Figs.~\ref{dataset-accuracy} and~\ref{dataset-inf} illustrate representative distributions of accuracy and inference time. The bimodal accuracy distribution reflects the coexistence of exact-match tasks (e.g., \texttt{MMLU}) and semantic-similarity evaluations (e.g., \texttt{TruthfulQA}), while the broad range of log-scaled inference times highlights the impact of model size, quantization level, and hardware heterogeneity. These variations motivate the need for orchestration mechanisms that jointly select model, quantization, and execution placement under resource and latency constraints.
%%%%%%%%%%%%%
%%%%%%%%%%%%%%%

We model accuracy prediction as a binary classification task, indicating whether a model produces a correct answer for a given query. As shown in Fig.~\ref{dataset-accuracy}, most datasets naturally provide binary correctness labels, while \texttt{TruthfulQA} yields continuous semantic-similarity scores. To obtain a unified target variable, \texttt{TruthfulQA} scores above 0.7 are mapped to correct predictions and the remaining scores to incorrect ones. In contrast, latency, CPU/GPU utilization, memory consumption, and response size are modeled as regression tasks. We evaluate lightweight predictors, including Random Forest (RF) and XGBoost (XGB), selecting the best-performing model for each prediction task.

We implemented \texttt{LMEdge} in \texttt{Python}~3.12.0 using the \texttt{PuLP} library~\cite{PuLP} and the \texttt{CPLEX} solver for both the heuristic scheduler and BILP formulation.
To emulate realistic network conditions, the two Kubernetes clusters replay measured \texttt{4G LTE} bandwidth traces~\cite{raca2018beyond} using \texttt{tc/netem}. Distinct traces are assigned to each cluster, while time offsets across instances introduce heterogeneous and non-synchronous bandwidth dynamics. Resource utilization is monitored every \qty{10}{\second} using \texttt{Prometheus}, \texttt{cAdvisor}, and \texttt{PromQL}. Deployment, configuration, and metric collection are automated through \texttt{Ansible}, \texttt{Helm}, and custom shell scripts. Query arrivals follow a Poisson process with rates $\nu\in{0.0004,0.0008,0.001}$ requests/ms (0.4, 0.8, and 1 query/s per instance). Scheduling is performed in fixed \qty{10}{\second} epochs over both newly arrived and deferred queries. We set $\rho_i$ to the number of CPU cores or the empirically determined maximum number of concurrent GPU inference streams. Unless otherwise stated, $\lambda=0.5$ and $\theta=0.1$, balancing congestion awareness and accuracy preservation. Queries are dispatched to the selected model via REST-based HTTP APIs using JSON payloads, with each model employing its default tokenizer context window. 

We compare \texttt{LMEdge} against two per-query orchestration baselines: \textbf{Random}, which uniformly assigns queries across devices without considering resource availability, and \textbf{Load-aware}~\cite{jang2025edge}, which routes queries to the least utilized device based on real-time resource monitoring. Other approaches discussed in Section~\ref{sec:RelatedWork} operate at coarser granularities (e.g., batching or session-level scheduling) and are therefore not directly comparable.
%%%
\begin{figure}[!t]
\centering
\begin{subfigure}[b]{0.32\linewidth}
    \centering
    \includegraphics[width=\linewidth]{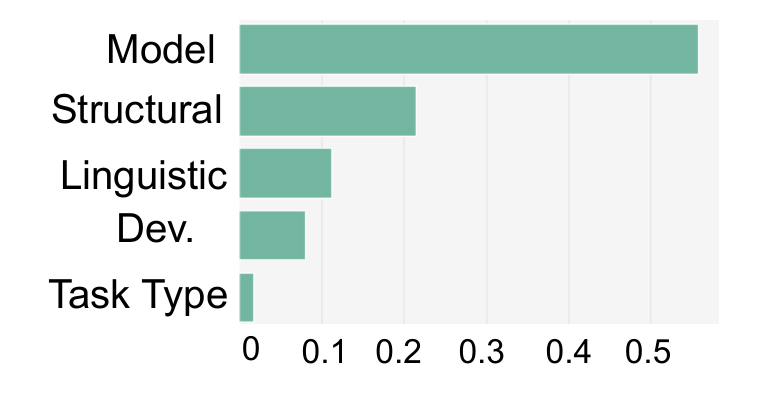}
    \caption{Inf. time.}
    \label{inf_imp}
\end{subfigure}
\hfill
\begin{subfigure}[b]{0.32\linewidth}
    \centering
    \includegraphics[width=\linewidth]{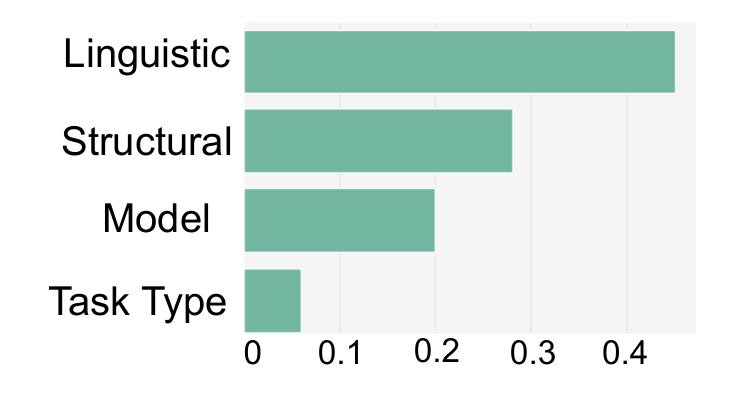}
    \caption{Acc.}
    \label{acc_imp}
\end{subfigure}
\hfill
\begin{subfigure}[b]{0.32\linewidth}
    \centering
    \includegraphics[width=\linewidth]{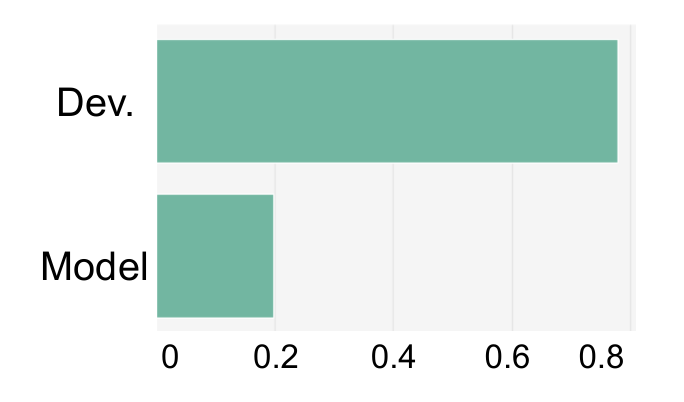}
    \caption{CPU/GPU.}
    \label{cpu_imp}
\end{subfigure}
\vspace{2mm}
\begin{subfigure}[b]{0.32\linewidth}
    \centering
    \includegraphics[width=\linewidth]{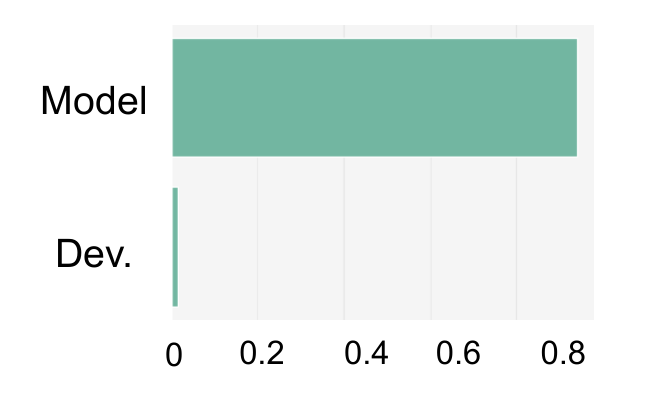}
    \caption{Mem. usage.}
    \label{mem_imp}
\end{subfigure}
\hfill
\begin{subfigure}[b]{0.32\linewidth}
    \centering
    \includegraphics[width=\linewidth]{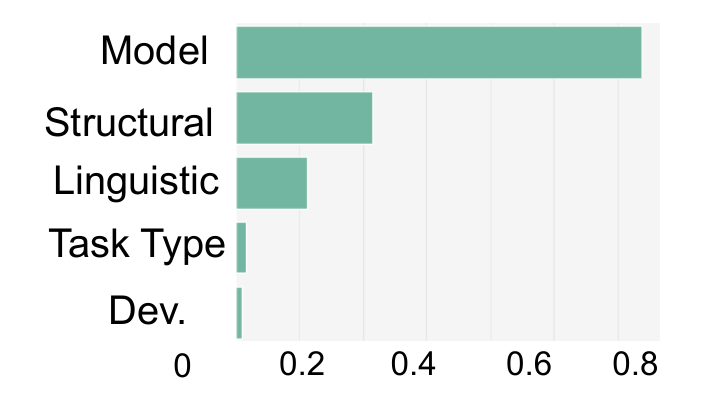}
    \caption{Res. size.}
    \label{resp_imp}
\end{subfigure}
\caption{Feature importance across the five prediction models.}
\vspace{-.3cm}
\label{fig:ML-predictors}
\vspace{-.3cm}
\end{figure}
%%%

%% file: 08_Experiments.tex
\vspace{-.3cm}
\section{Evaluation Results}
\label{sec:result}
\vspace{-.3cm}
We evaluate the performance of ML-based predictors using metrics reported in Tables~\ref{predictor-results}–\ref{predictor-results-acc} and Fig.~\ref{fig:ML-predictors}. XGB (\mbox{$n_{\text{trees}}=100$}, \mbox{$d_{\text{max}}=7$}, \mbox{$\psi=0.1$}) achieves the highest inference-time prediction accuracy, with $R^2=0.71$, $RMSE=0.59$, and $SDAE=0.14$, while requiring only \qty{0.07}{\second} for inference (\qty{3.9}{\micro\second} per query). As shown in Fig.~\ref{inf_imp}, model-related features are the strongest predictors of latency, followed by structural and linguistic query characteristics. The XGB classifier (\mbox{$n_{\text{trees}}=100$}, \mbox{$d_{\text{max}}=7$}, \mbox{$\psi=0.2$}) achieves the highest classification performance, with macro precision, recall, and F1-score of \num{0.83}, \num{0.79}, and \num{0.81}, respectively, while requiring only \qty{0.04}{\second} for inference. As shown in Fig.~\ref{acc_imp}, linguistic and structural query features are the strongest predictors of inference correctness, whereas model-related features have a comparatively smaller impact.
XGB (\mbox{$n_{\text{trees}}=100$}, \mbox{$d_{\text{max}}=5$}, \mbox{$\psi=0.3$}) accurately predicts CPU/GPU utilization, achieving $R^2=0.97$, $RMSE=177.3$, and $SDAE=0.06$ with negligible inference overhead. Fig.~\ref{cpu_imp} shows that device characteristics dominate resource consumption, while model-related features provide the second most influential contribution. XGB (\mbox{$n_{\text{trees}}=100$}, \mbox{$d_{\text{max}}=7$}, \mbox{$\psi=0.2$}) achieves highly accurate memory-usage prediction ($R^2=0.98$, $RMSE=206.6$, $SDAE=0.05$). As shown in Fig.~\ref{mem_imp}, memory consumption is primarily determined by model-related features, with quantization level emerging as the most influential factor. XGB (\mbox{$n_{\text{trees}}=200$}, \mbox{$d_{\text{max}}=7$}, \mbox{$\psi=0.1$}) achieves moderate response-size prediction accuracy ($R^2=0.61$, $RMSE=0.77$, $SDAE=0.12$) with minimal inference overhead (\qty{0.08}{\second}). This level of accuracy is sufficient, as response size plays a secondary role compared to latency and inference quality in scheduling decisions. As shown in Fig.~\ref{resp_imp}, response size is primarily influenced by model characteristics, followed by structural and linguistic query features, mirroring the trends observed for inference-time prediction.
%%%%%%%%%
\begin{table}[!t]
\centering

\caption{Performance of ML-based prediction models.}
\label{predictor-results}
\fontsize{7pt}{7pt}\selectfont
\begin{tabular}{|c|c|c|c|c|c|}
\hline
\textit{Prediction} & \textit{ML model} & \textit{$R^2$} & \textit{RMSE} & \textit{SDAE} & \textit{Inf. time} \\ \hline

\multirow{3}{*}{\textit{Inf. time}}
& \textit{XGB} & 0.71 & 0.59 & 0.14 & 0.07 \\ \cline{2-6}
& \textit{RF}  & 0.66 & 0.63 & 0.15 & 0.45 \\ \cline{2-6}
& \textit{MLP} & 0.66 & 0.64 & 0.15 & 0.13 \\ \hline

\multirow{3}{*}{\textit{CPU usage}}
& \textit{XGB} & 0.97 & 177.3  & 0.06 & 0.04 \\ \cline{2-6}
& \textit{RF}  & 0.97 & 177.2  & 0.06 & 0.10 \\ \cline{2-6}
& \textit{MLP} & 0.97 & 177.54 & 0.06 & 0.14 \\ \hline

\multirow{3}{*}{\textit{Mem. usage}}
& \textit{XGB} & 0.98 & 206.6  & 0.05 & 0.05 \\ \cline{2-6}
& \textit{RF}  & 0.98 & 207.6  & 0.06 & 0.12 \\ \cline{2-6}
& \textit{MLP} & 0.98 & 207.74 & 0.06 & 0.12 \\ \hline

\multirow{3}{*}{\textit{Resp. size}}
& \textit{XGB} & 0.61 & 0.77 & 0.12 & 0.08 \\ \cline{2-6}
& \textit{RF}  & 0.59 & 0.79 & 0.12 & 0.40 \\ \cline{2-6}
& \textit{MLP} & 0.61 & 0.79 & 0.10 & 0.20 \\ \hline

\end{tabular}

\vspace{-1mm}
\caption*{\textbf{Table 3.} Performance of ML-based prediction models.}

\begin{tabular}{|c|c|c|c|c|c|}
\hline
\textit{Prediction} & \textit{ML model} & \textit{Precision} & \textit{Recall} & \textit{F1-score} & \textit{Inf. time} \\ \hline

\multirow{3}{*}{\textit{Accuracy}}
& \textit{XGB} & 0.83 & 0.79 & 0.81 & 0.04 \\ \cline{2-6}
& \textit{RF}  & 0.81 & 0.78 & 0.79 & 0.17 \\ \cline{2-6}
& \textit{MLP} & 0.72 & 0.60 & 0.61 & 0.07 \\ \hline

\end{tabular}

\label{predictor-results-acc}

\end{table}
%%%%%%%%%%%%%%%%
%%%%%%%%%%%%%%%%%%%%%%%%%%%%%%%%%

We compare the BILP formulation and the \texttt{LMEdge} heuristic under query arrival rates $\nu\in{0.0004,0.0008,0.001}$. While the heuristic consistently executes within sub-second scheduling times, BILP runtime increases rapidly with workload intensity. The heuristic achieves speedups of up to \qty{1565}{\percent} at $\nu=0.0004$ and \qty{7012}{\percent} at $\nu=0.0008$, highlighting the limited scalability of exact optimization under realistic traffic conditions. Thus, the remainder of this section reports heuristic-based results when comparing \texttt{LMEdge} against the baseline schemes.

\begin{figure*}[!t]
\centering
\begin{subfigure}[b]{0.4\linewidth}
    \centering
    \includegraphics[width=\linewidth]{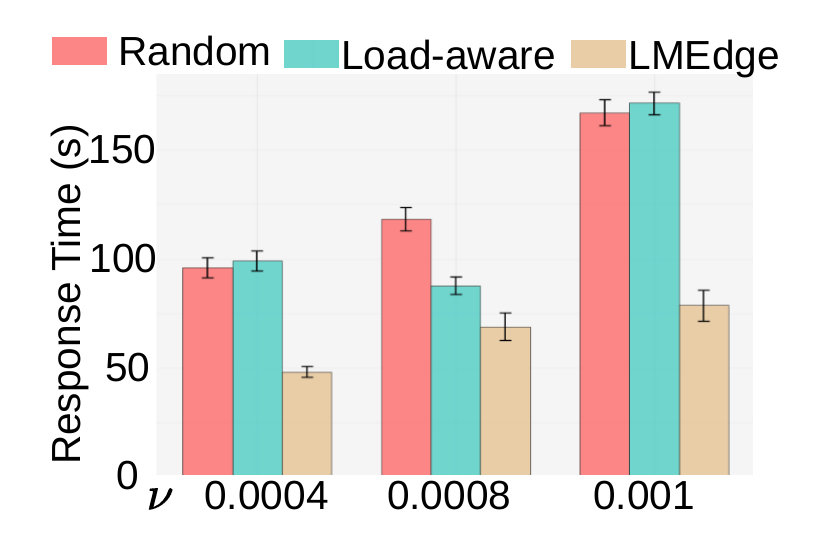}
    \caption{Response time.}
    \label{fig:baselines-poasson-time}
\end{subfigure}
\hfill
\begin{subfigure}[b]{0.4\linewidth}
    \centering
    \includegraphics[width=\linewidth]{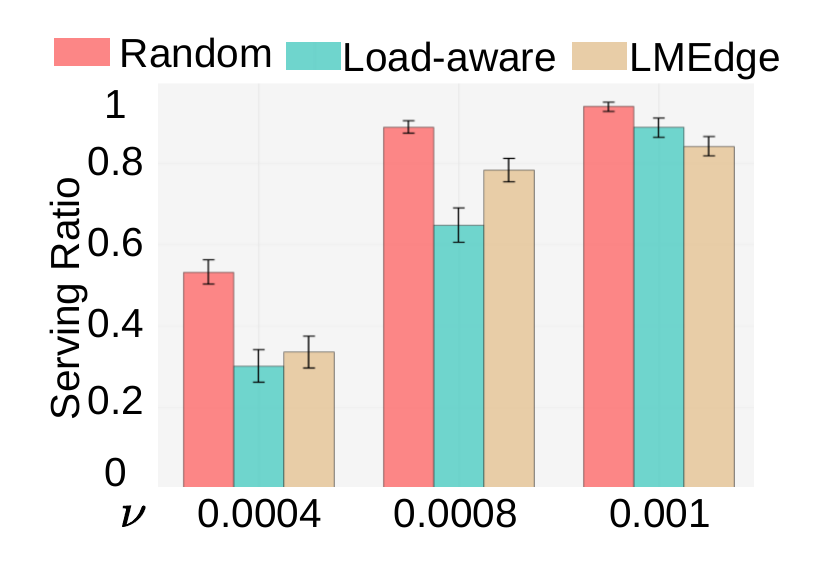}
    \caption{Serving ratio.}
    \label{fig:baselines-poasson-serv}
\end{subfigure}
\vspace{2mm}
\begin{subfigure}[b]{0.4\linewidth}
    \centering
    \includegraphics[width=\linewidth]{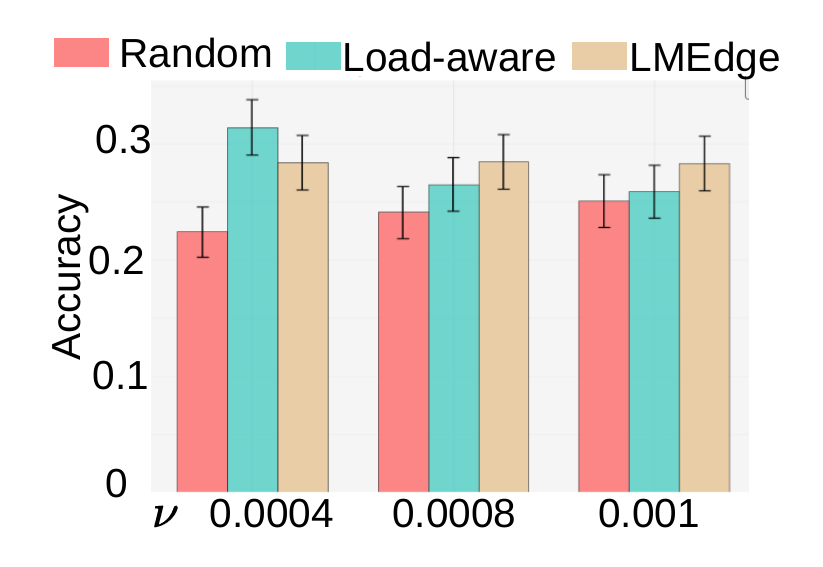}
    \caption{Accuracy.}
    \label{fig:baselines-poasson-acc}
\end{subfigure}
\hfill
\begin{subfigure}[b]{0.4\linewidth}
    \centering
    \includegraphics[width=\linewidth]{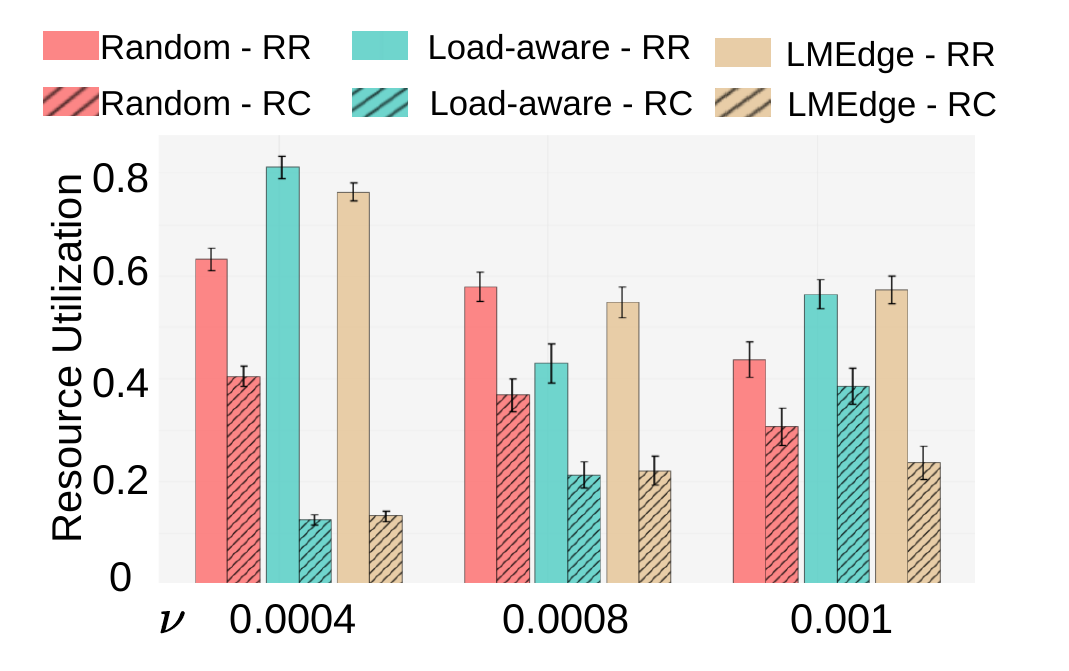}
    \caption{Resource utilization.}
    \label{fig:baselines-poasson-util}
\end{subfigure}
\vspace{-.5cm}
\caption{Performance of LMEdge vs. baselines under different query arrival rates.}
\label{fig:res}
\vspace{-.5cm}
\end{figure*}

\begin{figure}[!t]
    \centering

    \begin{subfigure}[b]{0.48\linewidth}
        \centering
        \includegraphics[width=\linewidth]{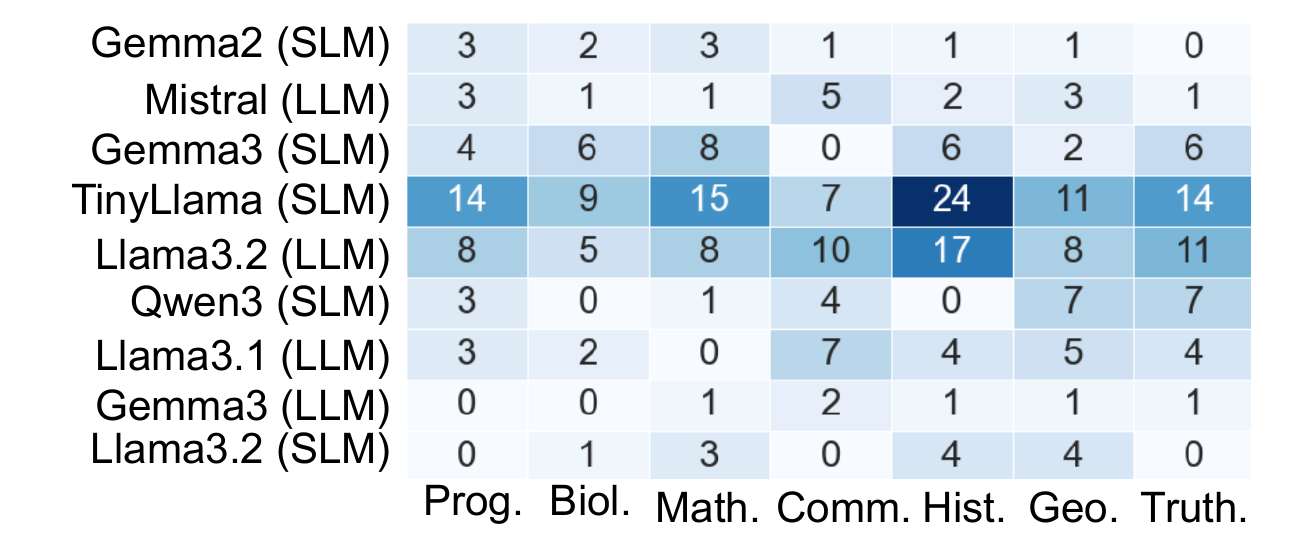}
        \caption{Random.}
        \label{fig:baselines-poasson-heatmap-rand}
    \end{subfigure}
    \hfill
    \begin{subfigure}[b]{0.48\linewidth}
        \centering
        \includegraphics[width=\linewidth]{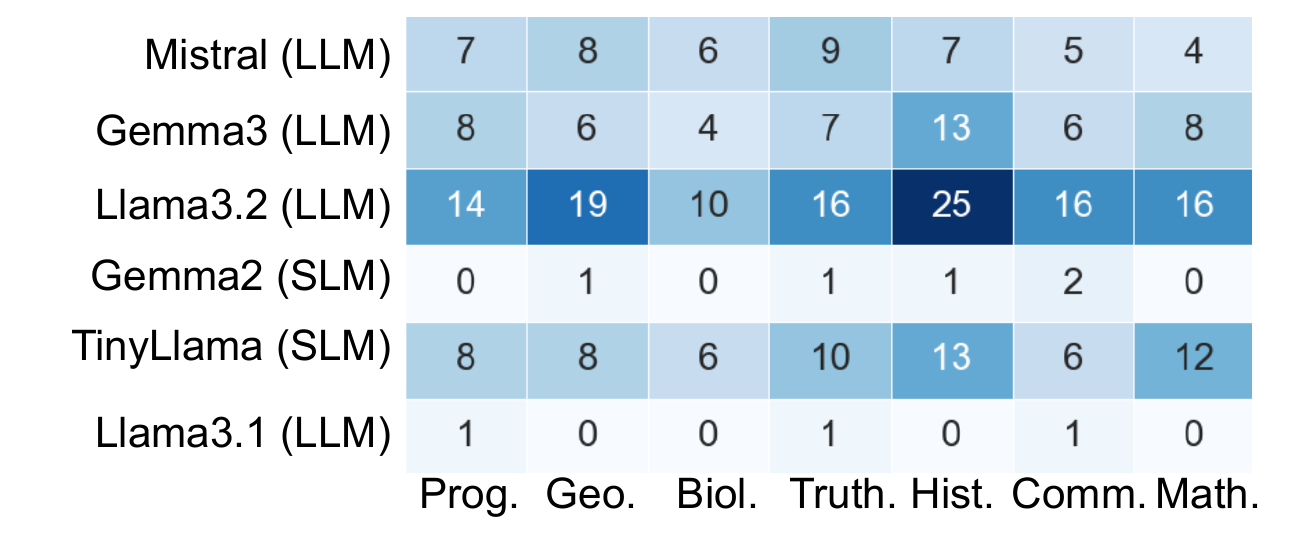}
        \caption{Load-aware.}
        \label{fig:baselines-poasson-heatmap-loadaware}
    \end{subfigure}
    \hfill
    \begin{subfigure}[b]{0.48\linewidth}
        \centering
        \includegraphics[width=\linewidth]{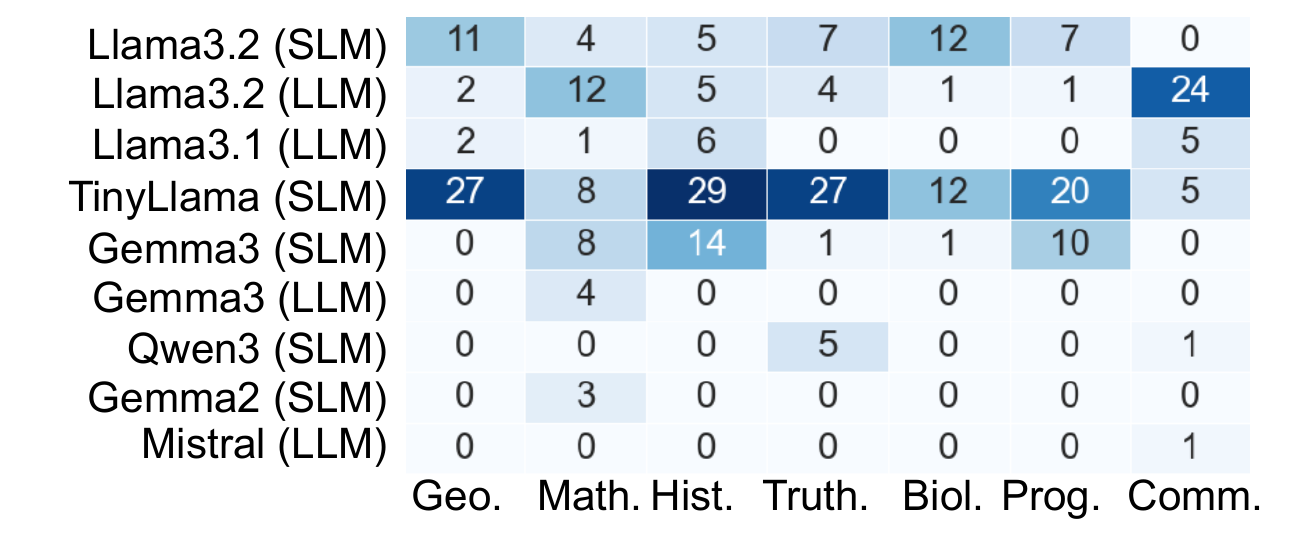}
        \caption{LMEdge.}
        \label{fig:baselines-poasson-heatmap-lmorch}
    \end{subfigure}
    \vspace{-.3cm}
    \caption{Query distribution across model configurations in different approaches.}
    \label{fig:baselines-poasson-heatmap}
    \vspace{-.5cm}
\end{figure}

Fig.~\ref{fig:baselines-poasson-time} shows that \texttt{LMEdge} consistently achieves the lowest response times across all arrival rates. Unlike Load-aware, which considers only instantaneous resource availability, and Random, which ignores both resource and QoS constraints, \texttt{LMEdge} jointly optimizes model, quantization, and placement decisions.
As shown in Fig.~\ref{fig:baselines-poasson-serv}, Random attains the highest serving ratio by processing all queries regardless of constraints, whereas \texttt{LMEdge} deliberately defers queries that violate accuracy or congestion requirements. Nevertheless, \texttt{LMEdge} consistently outperforms Load-aware and approaches Random's throughput at higher arrival rates. Fig.~\ref{fig:baselines-poasson-acc} further demonstrates that \texttt{LMEdge} maintains stable accuracy through the strict tolerance parameter $\theta=0.1$, while Random and Load-aware exhibit larger fluctuations under increasing load.
Fig.~\ref{fig:baselines-poasson-util} reveals distinct resource-utilization patterns. Random under-utilizes the available resources, whereas Load-aware concentrates workload on ERR resources and engages ERC devices only after saturation. In contrast, \texttt{LMEdge} balances workload across both layers, maintaining high ERR utilization while steadily exploiting ERC resources.
The heatmaps in Fig.~\ref{fig:baselines-poasson-heatmap} explain these trends. Random produces scattered assignments that ignore query characteristics, while Load-aware disproportionately favors large LLMs, leading to ERR congestion. \texttt{LMEdge} instead matches workload complexity to model capabilities, assigning lightweight LLMs to simpler queries and reserving larger LLMs for accuracy-sensitive tasks such as \texttt{TruthfulQA} and mathematics. This balanced use of LLMs reduces latency, improves resource utilization, and preserves accuracy across the edge devices.
%%%
%%%%%%%%%%%%%%%%
%%%%%%%%%%%%%%%%%%%%%%%%%%%%%%%%%

%% file: 09_Conclusion.tex
\vspace{-.3cm}
\section{Conclusion}
\label{sec:conc}
\vspace{-.3cm}
This paper introduced \texttt{LMEdge}, an orchestration service for serving LLM queries for edge infrastructures. We developed five ML-based predictive models, formulated the problem as a BILP optimization model, and proposed a lightweight online heuristic for efficient per-query orchestration. We constructed a benchmarking dataset with over \num{59000} rows covering diverse queries, models, quantizations, and devices, supporting reproducibility. Experiments on a realistic setup with \num{57} instances and \num{29} LLM configurations demonstrate that \texttt{LMEdge} outperforms baseline methods in terms of response time, accuracy, serving ratio, and resource utilization. Future work will investigate online learning, energy-aware scheduling, and larger GPU-equipped edge clusters.